\documentstyle[11pt,aaspp4]{article}

\def\bgc{$B_{gc}$}
\def\Mr{M^*}
\def\mpc17{Mpc$^{1.8}$}
\def\gtapr {\lower .1ex\hbox{\rlap{\raise .6ex\hbox{\hskip .3ex
        {\ifmmode{\scriptscriptstyle >}\else
                {$\scriptscriptstyle >$}\fi}}}
        \kern -.4ex{\ifmmode{\scriptscriptstyle \sim}\else
                {$\scriptscriptstyle\sim$}\fi}}}
\def\ltapr {\lower .1ex\hbox{\rlap{\raise .6ex\hbox{\hskip .3ex
        {\ifmmode{\scriptscriptstyle <}\else    
                {$\scriptscriptstyle <$}\fi}}}
        \kern -.4ex{\ifmmode{\scriptscriptstyle \sim}\else
                {$\scriptscriptstyle\sim$}\fi}}}
\begin{document}
\title{A Quantitative Measure of the Richness of Galaxy Clusters}
\author{
H.~K.~C.~Yee\altaffilmark{1,2}, and
Omar L\'opez-Cruz\altaffilmark{1,2,3}
}

\altaffiltext{1} {Department of Astronomy, University of Toronto, Toronto, 
Ontario M5S 3H8, Canada, Email: hyee@astro.utoronto.ca}
\altaffiltext{2}{Visiting Astronomer, Kitt Peak National Observatory,
KPNO is operated by AURA, Inc. \ under contract to the National Science
Foundation.}  
\altaffiltext{3}{Instituto Nacional de Astrofis{\'\i}ca, Optica y
Electr\'oonica (INAOE), Tonantzintla, Pue.,  M\'exico, Email: omar@inaoep.mx}

\received{}
\accepted{}

\begin{abstract}

Using photometric catalogs from wide-field CCD images, 
we derive the cluster-galaxy correlation
amplitudes, \bgc, for 47 low-redshift Abell clusters. 
We apply a number of tests to establish the robustness of 
the \bgc~parameter as a quantitative measure of richness of
galaxy clusters.
These include using different galaxy luminosity functions
to normalize the excess galaxy counts, counting galaxies
to different absolute magnitude limits, and counting galaxies
to different cluster-centric radii.
These tests show that with a properly normalized luminosity
function, the \bgc~parameter is relatively insensitive (at better
than 1/2 of an Abell Richness class)
to magnitude limit, areal coverage, and photometric errors
of up to about 0.25 mag.
We compare the \bgc~values to both the Abell Richness Class (ARC)
and Abell count numbers ($N_A$).
It is found that there is a good correlation between \bgc~and $N_A$
for Abell clusters with $z\ltapr0.1$, with a dispersion of about
one ARC; whereas the Abell Richness
classifications for Abell clusters at $z\gtapr0.1$ are much
less well correlated with the true cluster richness.
We also find evidence that the richness of ARC$\ge$3 clusters  has
the tendency of being over estimated in the Abell catalog. 

\bigskip
\noindent
{\it To appear in The Astronomical Journal, May 1999}
\end{abstract}

\section{Introduction}
The richness is an important defining
attribute of a galaxy cluster, yet has always proven to
be difficult to quantify rigorously.
Abell (1958), in creating the first comprehensive catalog
of rich galaxy clusters, defined the richness classification which
now bears his name based on the number of galaxies  within 2
mag of the third ranking galaxy in an angular area of radius 1.7\arcmin/$z$,
or about 3 Mpc (for $H_0=50$ km s $^{-1}$ Mpc $^{-1}$).
The procedure that Abell used in counting the number of galaxies
from Palomar Sky Survey Prints must be considered approximate,
and difficult to extend to higher redshift.
The accuracy of the Abell classification for Abell clusters
is accurate probably to only about 1 class (e.g., see Dressler 1978).
Uncertainties can arise from many sources: inaccurate estimate
of photometry, poor star-galaxy classification,
 incorrect background correction, contamination
from foreground or background clusters, and large uncertainty in
the estimated redshift.
In addition, the relatively large area used for the counting
increases the chance of contaminating projections.
At higher redshift, most of these problems would only be magnified, 
making a similar richness classification impractical without additional
data such as accurate photometry and redshift.
Furthermore, at redshifts higher than $\sim$0.2, 
it becomes increasingly more difficult, if not impossible, to specify, 
for example, the third brightest cluster member, due to the
large number of foreground galaxies.

A robust richness measurement for galaxy clusters is an important
observational quantity.
Beside being one of the primary attributes of a cluster,
it can be used as a baseline for comparing other properties 
of clusters in a meaningful way.
This is especially important as advances in technology have allowed
us to study galaxy clusters at different epochs and at increasingly
higher redshifts, up to and above 1 (e.g., Dickinson 1997).
Ultimately, a more fundamental parameter is the mass of the cluster.
However, there is evidence that the richness of a cluster, properly
determined, can be  
a good predictor of the mass of the cluster (Carlberg et al.~1996,
and Yee et al.~1999) based on results from
the CNOC1 cluster redshift survey.

Various attempts have been carried out to produce a more accurate
measure of the richness of Abell clusters. 
These include the $N_{0.5}$ of Bahcall (1981), and the revised
Abell Count Numbers ,$N_A$, of Abell, Corwin, \& Olowin (1989, hereafter ACO).
To arrive at a robust and rigorous quantitative measure of cluster richness,
three observational issues must be taken into account when counting
the number of galaxies in a cluster: 
the angular coverage, the depth of the observation, 
and the background correction.
The availability of wide field digital imaging due to advances
in large format CCDs has enabled us to address these issues
quantitatively much better.
The wide field allows us to sample the bulk of even very low redshift
clusters, and  digital imaging provides us with accurate photometry,
which is vital for determining the depth of the galaxy counts,
and correcting for background.
In addition, a useful quantitative
richness measure is one that can be applied uniformly to different
samples at vastly different redshifts.

The cluster-center galaxy correlation amplitude (\bgc), first
used by Longair \& Seldner (1979) to quantify the environment
of radio galaxies, is one parameter that
appears to be well-suited as a robust richness parameter.
By assuming an universal luminosity function (LF) and spatial
profile, the excess galaxy counts around any reference point
can be properly corrected to create an unbiased estimator
 of the richness of the environment.
Yee \& Green (1987) extended the method to study the environment
of quasars by using a ``universal'' LF with evolution that is self-consistent
with the background galaxy counts.
The ability to track the evolution of the LF at different epochs
is extremely important in comparing richness of clusters at different
redshifts in a meaningful way.
However, although such a method has been used in many studies of
the environments of active objects, fully digital photometric data
have not been applied to the quantification and calibration of the
richness of ordinary galaxy clusters, such as the Abell clusters,
and to test the robustness of the parameter.

In this paper, we present the richness measurements, in the form
of \bgc, for a sample
of 47 Abell clusters based on digital photometric data.
These measurements 
can be used as a calibration between the \bgc~parameter and
the more traditional Abell richness class (hereafter, ARC).
We show that \bgc~is a robust richness estimator, provided that
a  galaxy LF well matched to
that found in clusters is used for normalization.
We find that there is a good correlation between
the $N_A$ and \bgc~for clusters at $z\ltapr0.1$ with  a dispersion 
of about 1 richness class.
However,  $z>0.1$, $N_A$ often over-estimates the
true richness of Abell clusters.

  In \S2 we present the sample and the data set.
The method for deriving \bgc~is briefly discussed in \S3, along with
the LFs used in the computation.
In \S4 we test the robustness of the \bgc~parameter by investigating
the dependence of the parameter on various quantities such as
the LF, magnitude limit, and sampling area.
Section 5 discusses the results and compare the ARC
with the \bgc~parameter. 
A summary is presented in \S6.
Throughout this paper, unless otherwise specified, we use $H_0=50$
km s$^{-1}$ Mpc $^{-1}$ and $q_0=0.1$.

\section{The Cluster Sample and Data} 

The data used in this work are from 
a sample of Abell clusters originally chosen for a study
of correlations between X-ray and optical properties
(the Low-Redshift Cluster Optical Survey (LOCOS), see L\'opez-Cruz \&
Yee 1999).	
The detailed description of the sample, observation, and
data reduction can be found in L\'opez-Cruz (1997, hereafter LC97).
Here, we present a brief summary.

The clusters were
selected from a compilation of bright X-ray clusters
made by Jones \& Foreman (1998) with the following
criteria: (1) galactic latitude $|b|\ge30^{\rm o}$;
(2) $0.04<z<0.20$; (3) ARC$\ge$1; and (4) the declination
$\delta\ge24^{\rm o}$.
Although there is an imposed criterion of ARC$\ge$1, data for
a number of ARC=0 clusters were also obtained, often for the
purpose of filling Right Ascension gaps during the observing
runs.
The ARC=0 clusters are not considered as part of the fair sample.
This is because they were not picked randomly;
ones that appear to be rich or have well-defined X-ray emission
were preferentially chosen.
Nevertheless, they are analysed along with the other clusters, but are
used only when such a selection effect does not affect
the particular conclusion.
As a comparison, the central region of Coma (A1656, $z=0.0232$) was
also observed.
The total number of Abell clusters in the sample is 47,
with 36 satisfying the criteria listed above.
The cluster names, along with their redshifts and Abell Richness Classes, 
are listed in Table 1.

 Direct images in $B$, $R$, and $I$ were obtained using the
T2KA CCD camera at the 0.9m telescope at KPNO over five
runs in 1992 and 1993.
Photometry was calibrated to the Johnson-Kron-Cousin system
of Landolt (1992).
The CCD has 2048$\times$2048 pixels with a scale of 0.68$''$
pixel$^{-1}$, giving a field of view of 23.2$'\times$23.2$'$.
The integration times varied between 900 to 2500 second,
depending on the filter used and the redshift of the cluster.

Nine control fields at least $5^{\rm o}$  away from any cluster fields
were also observed for the purpose of background count correction.
These fields were observed and  reduced in an identical manner as 
the cluster fields.

Photometric catalogs of all objects in each field were created using
the program PPP (Yee 1991) to perform automatic object
finding, photometry, and galaxy-star classification.
The typical 100\% completeness magnitude in $R$ is $\sim21.5$ mag.
This provides a range of completeness in absolute $R$  magnitudes at 
the redshift of the clusters from --15.5 to --18.5.

\section{Analysis}
\subsection{The \bgc~Parameter}

The cluster richness parameter \bgc~is defined as the amplitude
of the cluster center--galaxy correlation function:

$$ \xi(r)=B_{gc} r ^{-\gamma}. \eqno(1)$$

This parameter was first used by Longair \& Seldner (1979)
to quantify the richness of radio galaxy environment.
It was subsequently adopted by Yee \& Green (1984) to study
the environment of quasars using a self-consistent 
LF estimated from the same data.
Further applications of the parameter to the study of the environments
of quasars and radio galaxies can be found in Yee \& Green (1987; hereafter
 YG87); Ellingson, Yee, \&
Green (1991); Prestage \& Peacock (1988); and Yates, Miller, \& Peacock
 (1989).
This method has also been used to quantify the environment of
BL Lacertae objects (Smith, O'Dea, \& Baum 1995;
Wurtz et al. 1997), Seyfert galaxies
(De Robertis, Yee, \& Hayhoe 1998), and the richness of a
 small sample of Abell clusters (Anderson \& Owen 1994).
Detailed descriptions of the method adopted for this study
can be found in YG87, 
and also Andersen \& Owen (1994), among others.
Here, we provide a brief outline.

We observe galaxies in projection on the sky, which allows us to
measure the angular two-point correlation function of galaxies, 
$\omega(\theta)$,
as a function of the angle $\theta$ on the sky. 
The function $\omega(\theta)$ can be approximated by a power-law of 
the form (e.g., Davis \& Peebles 1983):

$$ \omega(\theta)=A_{gg} \theta ^{1-\gamma}.\eqno(2)$$

\noindent
where $A_{gg}$ is the galaxy-galaxy angular correlation amplitude.
For our purpose,
we can interpret $\omega(\theta)$ as the distribution of
galaxies projected on the sky around a reference point, such
as the center of a galaxy cluster.
In this case, the amplitude, now notated as $A_{gc}$,
can be measured directly from an image by counting
the excess (i.e., background corrected) galaxy counts, $N_{net}$,
up to a certain apparent magnitude
around this reference point within some $\theta$;
i.e., if a fixed $\gamma$ is assumed, then
$ A_{gc}={N_{net}\over{N_{bg}}} {(3-\gamma)\over{2}} \theta^{\gamma-1}$,
where $N_{bg}$ is the background counts within $\theta$.

The \bgc~amplitude can then be estimated via a 
deprojection of the angular correlation function into
the spatial correlation function by assuming spherical symmetry, 
giving the relation between \bgc~and $A_{gc}$ (see Longair \& Seldner
1979):

$$ B_{gc}=N_{bg} {D^{\gamma-3} {A_{gc}}\over{I_\gamma
\Psi[M(m_0,z)]}},\eqno(3)$$

\noindent
where $N_{bg}$ is the background galaxy counts to apparent magnitude
$m_0$;
$\Psi[M(m_0,z)]$ is the integrated LF of galaxies up to
the absolute magnitude $M$ corresponding to $m_0$ at the redshift of the
cluster $z$.
$I_\gamma$ is an integration constant arising from the deprojection,
(with $I_\gamma=3.78$ for $\gamma=1.77$), and $D={c \over H_0} \{q_0z+(q_0-1)
[(2q_0z+1)^{1\over 2}-1]/q_0^2(1+z)^2\}$ is the angular diameter
 distance to $z$.

As eqn (3) indicates, the computation of \bgc~requires 
a knowledge of the luminosity
function (which provides a normalization) and a background 
count correction. These are described in the next two subsections.

\subsection {The Control Field Counts}

The background counts used were obtained from the same cluster
imaging program using the same telescope-CCD-filter combination
and the same reduction procedure, which are described in detail
in LC97.
Briefly, 9 fields with integration times similar to those of
the cluster fields
were obtained at least 5$^{\rm o}$~from any of the program clusters.
The control field counts were combined using the technique described
in Yee, Green, \& Stockman (1986).
Figure 1 shows the counts.
These counts are in good agreement with the red band counts from
other investigations (e.g., Driver et al. 1994, Metcalfe et al. 1991,
and Yee et al. 1986).  Using background counts obtained in an
identical manner as the cluster data removes one of the major
systematic uncertainties in the determination of \bgc.

\subsection {The Luminosity Function}

The choice of the LF used is a crucial one in the determination of \bgc. 
The sensitivity of the \bgc~value to the LF will be discussed in detail
in \S 4.1.
The assumption of a universal LF is not strictly correct, as
field and cluster galaxies have different population mixes and
evolution histories.
However, for the computation of \bgc~using galaxies brighter than
$\sim -18$ mag, the varying faint-end slope of cluster galaxy
LF (L\'opez-Cruz et al.~1997) has little or no effect.
We choose to test three LFs derived from very different data sources
and methods. They are all represented functionally
by the Schechter function (Schechter 1976).
The three functions are (1) the LF derived from the clusters themselves
(L\'opez-Cruz \& Yee 1999, hereafter the LCY LF);
(2) the low-redshift field galaxy
LF from King \& Ellis (1985, hereafter the KE LF); and
(3) the field galaxy LF from the CNOC2 redshift survey 
of intermediate redshift galaxies (Lin et al. 1999,
hereafter the CNOC2 LF).
The parameters of the LFs are listed in Table 2 and the LFs are
described in the following.
We will first discuss primarily the shape of the LFs used and 
defer the description of the normalization of the LFs
to later in the section.

We derive the shape of the LCY LF using the cluster data themselves.
The derivation of the LF for this sample of clusters
is described in detail in LC97,
L\'opez-Cruz et al. (1997), and L\'opez-Cruz \& Yee (1999).
The bright end is fitted to a Schechter function with  
$\alpha=-1.0$.
The result (LC97, L\'opez-Cruz \& Yee 1999) shows that the 
sample has $<M^*>=-22.26
\pm 0.29$, where the uncertainty is the dispersion of the distribution.
However, cD clusters appear to have a larger range of $M^*$ which is
apparently correlated with richness (LC97, L\'opez-Cruz \& Yee 1999). 
Removing the cD clusters produces a $<M^*>=-22.32\pm0.26$.
We adopt an ``universal'' LF for the clusters with $M^*=-22.3$ (at
a median redshift of 0.07) and $\alpha=-1.0$.

For deriving richness measurements for clusters at significant
redshifts, the evolution of galaxies comes into play.
Counting galaxies to a fixed absolute magnitude will in general
over estimate the richness for high redshift clusters, as even
mild passive evolution will mean effectively counting deeper into the LF.
Hence, following YG87, we add a mild luminosity
evolutionary term parametrized as  $-Qz$ mag, where $Q\sim1.4$,
to $M^*$ to compensate for this effect.
(YG87 used a second order polynominal for $M^*$ evolution, we have
simplified this to a single term in $z$, since any such measurement is
currently extremely approximate.)
There is good evidence that the LF in clusters evolves, to first order,
in such a manner from galaxies associated with quasars (YG87),
and from the CNOC1 cluster redshift survey (Yee et al. 1999).
Since the median redshift for the LC97 clusters is 0.07, we correct
the $M^*$ to redshift 0 to $\sim 22.2$.

For a consistency check with previous applications of the $B$ parameters in 
analyses of the environments of quasars (e.g., YG97), radio galaxies
(Yates et al. 1989), BL Lacs (Wurtz et al.~1997), and Seyfert galaxies
(De Robertis et al. 1998), we also compare \bgc~values 
using the field galaxy KE LF with its separate
components for different morphological types.
The KE LF was derived as a preliminary result
using a small number of low-redshift 
galaxies from the AAT/Durham redshift survey.
Here, the $M^*$ values are transformed into the Cousin $R$ band
from the Gunn $r$ band as listed in YG87.
The same luminosity evolution as that for the LCY LF is
incorporated into the KE LF.

The final LF used is the interim LF derived from the CNOC2 field
galaxy redshift survey (Yee et al.~1997, Lin et al.~1999).
The LF is determined based on about 2300 galaxies with redshifts
between 0.1 and 0.55, covering two patches on the sky.
The galaxies are separated into early, intermediate, and late
spectral types based on 5-color photometry.
The LF for each component is then fitted to a Schechter function
with two evolution parameters, $P$ and $Q$, in the form of
$\Phi(z,M)=\Phi(0,M-Qz)10^{0.4Pz}$ (see Lin et al.~1999 for details).
Hence, the evolution of this LF is derived explicitly with
$Q$ measuring the luminosity evolution, and $P$, the density
evolution (see Table 2).

An important parameter in the LF for the determination of \bgc~is
 the normalization constant $\phi^*$.
The value of \bgc~is inversely proportional to it.
As discussed in YG87, the best way to minimize systematic
effects in \bgc~is to  determine a self-consistent LF with a
normalization constant that reproduces the background counts as 
a function of apparent magnitude.
We hence renormalized $\phi^*$ by multiplying it by $\phi'$ which
is determined by fitting the modeled  counts to the background counts.
We determine $\phi^*$ for the LCY LF directly by minimizing
the $\chi^2$ between model counts generated by using the LF and
the background count data.
For the KE and CNOC2 LFs, the relative $\phi^*$'s for each
component are retained, but an overall normalization is applied
based on the background count fitting.
We found that the renormalization of the CNOC2 LF, which is derived using
the same photometry band, is small, less than 10\%.
The modeled counts for the various LFs are shown in Figure 1.

\subsection {Results and Error Estimate}

The \bgc~values obtained using the LCY LF are listed in Table 1.
The values obtained using the other two LFs are similar, with variations 
of less than $\sim$10\%.
A more detailed comparison of the dependence of \bgc~on the LF is
left to Section 4.1.
For each cluster, we count galaxies down to a limit of $M_R=-20$
 in a circle of 0.5 Mpc radius nominally 
using the brightest cluster galaxy (BCG) as the center of the cluster.
However, we note that a few clusters have their BCG not sitting in the
obvious center of the distribution of galaxies.
The most significant deviation occurs in A168, where the BCG is almost
isolated from the cluster.
In this case, we have chosen the brightest galaxy in the high density
clump (with the correct color) to be the center of the cluster.
Choosing the BCG as the center for A168 would produce a \bgc~almost 3 times
smaller.

  The uncertainty for the \bgc~parameter is computed using the
formula:

$$ {\Delta B_{gc}\over B_{gc}} = {{(N_{net}+1.3^2N_{bg})^{1/2}}
\over {N_{net}}}. \eqno(4)$$

\noindent
where $N_{net}$ is the net counts of galaxies over the background of
$N_{bg}$.
This is a conservatively large error estimate as it includes 
the expected counting statistics in $N_{net}$ and the expected
dispersion in background counts.
The factor $1.3^2$ is included to account approximately for the additional
fluctuation from  the clustered (and hence non-Poissonian) nature
of the background counts (see LC97 and YG87).
This is a fair estimate of the uncertainty of the
richness of the cluster, as it
includes the Poissonian uncertainty in the net counts, which
is equivalent to the uncertainty in the number of galaxies
drawn from a Schechter function with a fixed normalization (hence of
fixed richness).
However, we note that the inverse of this  
error estimate is an underestimate of
the statistical significance of the excess galaxy counts in the sample,
which would be better represented by 
$N_{net}/1.3(N_{bg})^{1/2}$.
Hence, at these low redshifts, while the average
uncertainty in the \bgc~parameter
is about 25\%, the significance of the excess counts over background
is considerably better at about the 20$\sigma$ level.

\section{The Robustness of the \bgc~Parameter}

In this section, we examine the results of
exhaustive tests on the dependence
of \bgc~on various quantities, such as the LF shape and parameters,
sampling and spatial limits, and photometric errors.
These tests not only provide a clear indication of the robustness of 
the \bgc~parameter, but also the pitfalls in applying the method.

\subsection {The Dependence of \bgc~on the Luminosity Function}

A richness measurement based on the number of galaxies in a cluster
is invariably tied to the assumption of the luminosity function
of the galaxies.
If clusters have greatly different galaxy LFs, then any such richness
measurement would be rendered meaningless.
In this section, we test the sensitivity of \bgc~to the 
LF used in converting $A_{gc}$ to $B_{gc}$.
For these tests we vary the LF parameters but fix the sampling limit 
at $M_R=-20$ mag. 
This sampling limit was chosen with the expectation that it minimizes
the effect of having an incorrect LF.
Varying the LF with different sampling limits causes more
complex behaviors; these effects are tested in \S 4.2.

\subsubsection {Comparing Results from Different Luminosity Functions}

We have computed \bgc~using three LFs derived using very different
data and method.
The LCY LF is based on the cluster LF measured from the same set of
data, and describes cluster galaxies the best;
while the KE and CNOC2 LF are based on field galaxies at different
redshifts.
The three give remarkably similar results with the mean of the ratio
of \bgc (LCY):\bgc(CNOC2):\bgc(KE) being 1.00:1.10:0.92, with the mean
value for \bgc(LCY) being 1246$\pm$417,
where the uncertainty is the dispersion of the \bgc~values in the
sample.
This result demonstrates that the systematic uncertainty arising from the
not-strictly-correct assumption of a ``universal'' LF for both
cluster and field galaxies is of the order of 10\%. 
The ratios indicate that the CNOC2 LF is slightly fainter overall
compared to the LCY LF, while the KE LF is  somewhat brighter.
This indeed is the case when comparing the $M^*$'s of early type galaxies
in the two multi-component LFs with that of the LCY LF.

For the remainder of the paper, we will use the \bgc~values
derived using the LCY LF as the fiducial set, and compare the results
from the various tests to it.

\subsubsection {The Effect of an Incorrect Luminosity Function}

First, we test the effect of an ``universal'' LF with
incorrect parameters.
In varying the $M^*$ and $\alpha$ of the LF, we also adjust the
normalization constant of the LF, so that it reproduces the
control field galaxy counts up to the magnitude of sampling.
Note that this is an important step to make sure that the \bgc's
are obtained with a self-consistent LF model.
This is different from testing the effects of statistical fluctuations
of the LF of individual clusters from the mean for which
one would not perform the renormalization (see \S 4.1.3).

In these tests we first adjust the LCY LF by altering $M^*$ in
steps of $\pm$0.25 mag. 
Each time we recompute the normalization
constant for the LF by fitting the background counts.
We note that the fit to the background counts deteriorates with
each step, but remains reasonable up to a $\Delta M$ of $\pm$0.5 mag.
The results are summarized in Table 3.
The results for $M^*$ being off by $\pm$0.25 mag are quite
acceptable, causing variations of about 10\%, similar to the
effects of using the three different LFs.
This is not surprising, as the uncertainties in the three LFs are
of the same order.
Even when the error is 0.5 mag, the effect is about 20\%, comparable
to about 1/2 an Abell Richness Class (equivalent to $\Delta B_{gc}$
of about 200; see Section 5).

We next alter the faint-end slope $\alpha$ of the LF, again, 
with renormalization based on fitting the background.
Here, we change $\alpha$ by steps of 0.15.
The results are tabulated in Table 4.
Again, one sees that changing the slope by as much as $\pm0.3$
alters the \bgc~values by less than 20\%.

These tests, along with the results from using three different LFs,
demonstrate that  a LF which self-consistently reproduces the
background counts can
be substantially incorrect and still produces results that
are accurate to better than 1/2 an Abell class.

\subsubsection {The Effect of Fluctuations in the Cluster Galaxy
Luminosity Function}

In this section, we test the effect of statistical fluctuations
from the mean in the LF of the cluster galaxies.
Here, we vary  $\Mr$  and $\alpha$,
but do not adjust the normalization constant of the LF.
Note that in the case of $\Mr$, it has the same effect as a 
photometric error of the same magnitude 
in the cluster data.

  We use the same step sizes for $M$ and $\alpha$ as in \S 4.1.2,
and the results are listed in Tables 3 and 4.
Here, we see that the effects are significantly larger.
Specifically, a fluctuation in $M^*$ (or equivalently,
 photometric error in the data)
of about 0.4 mag will produce a 20\% (1/2 Abell class) effect.
However, a fluctuation of 0.2 mag would still be quite acceptable,
producing uncertainties of only about 10\%.
Since cluster $M^*$'s have a dispersion of about 0.2 mag, this indicates
that the instrinsic dispersion of the LF from cluster to cluster does
not affect \bgc~significantly.
Similar fluctuations due to variations in $\alpha$~of up to 0.3 are
also seen.

\subsection {The Dependence of \bgc~on the Limiting Magnitude}

The independence of \bgc~values on galaxy counting limiting magnitude 
is an important issue, especially when different fields are 
observed to different depths.
If the correct LF is chosen for the computation of \bgc~and the
LF is truly universal, then \bgc~should subject
only to counting statistical variations when different limiting
magnitudes are used.
A substantial systematic variation in \bgc~with limiting magnitude can
be taken as an indication that a LF with the incorrect shape
has been chosen.
Moreover, for the purpose of producing robust results, there
is an optimal range of absolute magnitudes for galaxy counting.
Counting to too bright a limit will produce a large uncertainty
from counting statistics.
Furthermore, one is at the mercy of small intrinsic 
variations of $M^*$ in individual clusters, or equivalently,
small photometric zero point error in the galaxy photometry.

On the other hand, a larger uncertainty may also result
from counting too deep into the LF, as the background counts
rise more rapidly than the flat part of the cluster LF.
In addition, in counting beyond $M_R\sim-18$ one may encounter
 large variations in the faint end slopes of the cluster LF
(L\'opez-Cruz et al. 1997 and LCY), producing additional fluctuations.
Ideally, one would like the richness parameter of a galaxy cluster
to be representative of the total mass of the cluster, which,
under the light-traces-mass scenario (e.g., Tyson \& Fischer 1996;
 Carlberg, Yee, \& Ellingson
1997), should be proportional to the total light.
Hence, the richness of a cluster is best represented
by the number of $L^*$-like galaxies, rather than 
by the number of dwarf galaxies;
consequently, a robust measure of \bgc~is best obtained using a magnitude
limit corresponding to the flat part of the LF: from about 1 mag
to 3 mag past $M^*$.

  The data in our sample of Abell clusters sample to at least $M_R=-18.5$
mag for all the clusters; hence, they provide a good test on the 
robustness of \bgc~versus counting magnitude, 
and the effects of varying the LF.
For these tests,
we compute the \bgc~parameters using $M_R=-22.0$, --21.0, -20.0,
--19.0, and --18.0 for comparison
with the standard limit of --20 (i.e., about 2 mag past $M^*$).
The five derivations provide ratios of average \bgc~of:
1.04:1.05:1.00:1.00:1.07.
A comparison of the individual \bgc's using sampling limits of --21.0
and --20.00 is shown in Figure 2.
This test shows very strongly that, as expected,
when a LF closely resembling that
of the cluster galaxy LF is used, the sampling magnitude limit has
little or no systematic effect on the derived richness.
The variations in individual \bgc~values are  the order of the
computed uncertainties, and are due to counting statistics.
We note that the largest deviation occurs with the deepest sampling
limit.
This is expected as the steepening of the galaxy LF for some clusters
begins at about $M_R=-18$, which would produce a larger count relative
to the flat ($\alpha=1$) LF used.

Next, we consider the effects of the combination of LF variations
and more extreme sampling limits.
We focus on testing the effect of incorrect LF and photometric errors
when sampling to only about $M^*$, a situation that may arise often
when data which are not quite deep enough are pressed into use.
The results are shown in Table 3.
In general, at a sampling limit of $M_R=-22$ mag (about $M^*$),
the errors committed when $M^*$ is varied with renormalization
of the LF (equivalent to using an incorrect LF) are about twice as large
as those when the sampling is to $\sim$2 mag past $M^*$ (about --20 mag).
However, when no renormalization is performed, which is also equivalent
to photometric uncertainty, the errors incurred when counting to
--22 mag can be as large as 4 times those to --20 mag.
For example, a +0.5 mag change in $M^*$ produces \bgc~values 2.2 times
larger when counting to --22 mag, compared to 1.3 times larger when counting
is done to --20 mag.
We note that the results for changing $\alpha$ are not nearly as drastic, 
as that affects mainly the faint end of the LF (see Table 4).

We hence conclude that counting to about $M^*$ produces results that
are extremely sensitive to errors in the assumed LF, photometry, and
in general statistical fluctuation, with results that could be off by
a factor of 1.5 to over 2 when the error in magnitude is 0.25 to 0.5
mag.
This may explain the large discrepancies noted in the \bgc~values of
BL Lac objects between Smith, O'Dea, \& Baum (1995) and Wurtz et al. (1997), as
noted by Wurtz et al..
We also note that this may also be the possible explanation of the
discrepancies of between the richness derived for Abell clusters by
Andersen \& Owen (1994) and this work (see \S 5.1).

\subsection {The Dependence of \bgc~on Sampling Area}

Ideally, a robust richness parameter is one such that the counting
area used should not have a significant effect on the richness
derived.
Such a parameter would allow comparison of the richness parameter from
data taken with a wide variety of areal coverage.
This is strictly possible only if the spatial distribution
function of cluster galaxies has a similar form for all clusters.
If an incorrect slope ($\gamma$) for the correlation function is used, 
the \bgc~value will be dependent
on the sampling area.
In the computation of \bgc~it is assumed that the galaxy distribution
drops off in the same fashion as the spatial correlation function 
for field galaxies.
This assumption is adopted so that comparisons over a wide range
of richness and environments, including that of field galaxies,
can be made.
However, this assumption is most likely incorrect for rich clusters,
as various investigations have shown that the power-law slope 
of their galaxy density
distribution is probably greater than 2 (e.g., Lilje \& Efstathiou 1988,
Pebbles 1993),
somewhat steeper than that of the clustering of field galaxies.

The effect of using an inexact $\gamma$ can be examined by
comparing the \bgc~values for our sample using different limiting
radii.  We have computed \bgc~values (with the LCY LF)
using sampling radii of 0.25, 0.5, 1.0 Mpc, 
all with a $M_R$ sampling limit of --20 mag.
The ratio of the mean \bgc's for the 3 radii are: 1.03:1.00:0.94.
The results for individual clusters with sampling radii of 0.5 and 1.0 Mpc
are shown in Figure 3.

The results indicate that a change of a factor of 2 in the sampling radius
causes a scattering of about 5\% in the richness parameter.
This small effect is reassuring, and provides confidence when comparing
data sampled over metric field sizes of a factor of several, as in those
used by De Robertis et al.~(1998) for nearby Seyfert galaxies.
The decreasing trend in the \bgc~with larger radii is consistent with
existing evidence that cluster galaxy distribution is likely to be 
steeper than the slope of the field galaxy correlation function.
However, since the effect is very small,
we have chosen to retain the canonical $\gamma=1.8$ so that direct
comparisons with field and group clustering amplitudes can
be made.

\section{Discussion}

The various tests in Section 4 have established the robustness of the
\bgc~parameter as a measure of cluster richness.
In this section, we compare the \bgc~parameter with the Abell
Richness Class, establishing 
a calibration between the two.
We also discuss the correlation between \bgc~and the Abell
count number, which allows us to examine the accuracy of the Abell
count number as a measure of richness.
Based on these comparisons, we suggest a revised
 richness classification scheme based on the calibration between the
 \bgc~parameter and ARC.
Finally, we illustrate the advantage of using a robust
richness parameter in the investigation of cluster properties by
examining the correlation between richness and velocity dispersion.

\subsection {Comparison with Abell Richness Class}

One of the main goal of this investigation is to establish a calibration
between a more quantitative richness parameter, such as \bgc, and
the traditional Abell Richness Class.
In Figure 4, we plot the \bgc~parameters versus ARC.
We see in  general a very broad correlation between \bgc~and ARC.
We note that because we have specifically chosen to image
ARC 0 clusters which appear to be the richest, they should
not be considered in the correlation.

  Andersen \& Owen (1994) produced a qualitatively very similar
plot  of the two quantities using
a different sample and very different type of data.
However, quantitatively, there is a significant discrepancy in the
scaling of the two quantities:
Their \bgc~values are close to a factor of two lower than ours (after
correcting for $H_0$), with their average \bgc~for ARC 1 clusters
being $\sim 600$ Mpc$^{-1.8}$, compared to about 1000 Mpc$^{-1.8}$
in our work.
This discrepancy is perhaps not surprising, given that: (1) Andersen \&
Owen use POSS photographic data which may not be well calibrated,
(2) their object counting goes down to only $\sim M^*$, and
(3) they did not use a LF that has been renormalized to the background.
All these can contribute substantially to the systematic uncertainty,
up to a factor of 2, as demonstrated by our tests in \S 4.

Since the ARC is an approximate and quantized measure, we will next
look at the comparison with actual Abell counts ($N_A$) before
revisiting the correlation between \bgc~and ARC in \S 5.3.

\subsection {Comparison with Abell Count Numbers}

A more detailed comparison between the ARC and the \bgc~measurement
can be obtained by using the Abell number counts ($N_A$) as published
by ACO.
We plot $N_A$ vs \bgc~in Figure 5.
Taken as a whole, again we see a definite correlation between \bgc~and
$N_A$ but with a large dispersion.
Specifically, it is noted that at large $N_A$ the correlation
breaks down in that their corresponding \bgc~values are smaller than 
indicated by $N_A$.
We also note that all the clusters in our sample with large $N_A$s 
are those at higher redshifts.
We hence break the sample into two groups, dividing them at $z=0.09$.
To increase the $z>0.09$ sample, we also add 3 clusters from the
CNOC1 survey (Yee et al. 1999) which are Abell clusters at $z\sim0.2$.
These are A2390, MS0451+02 (A520), and MS0906+11 (A750).

Removing the high redshift sample provides a much improved correlation
bewteen \bgc~and $N_A$.
For the lower $z$ sample, though 
the scatter is about one Abell class, there are  few or no outliers.
For the higher $z$ sample, there is almost no correlation between
\bgc~and $N_A$.  
The discrepancies come entirely in the high $N_A$ clusters 
which all have too large a $N_A$ for their \bgc. 
It is interesting to note that A665, which is the only Abell cluster
classified as ARC 5,
with $N_A=321$, has a \bgc~value which indicates that its richness is
overestimated by a factor of 3 by $N_A$.
A number of Abell clusters in fact have similar richness to A665,
 including A2390 and A401.
Hence, we conclude that at $z\ltapr0.1$, the ACO counts provide reasonable 
estimates of the richness of the clusters; however, at $z\gtapr0.1$, and/or for 
$N_A\gtapr140$, the $N_A$ counts almost always over predict the cluster's
richness.
This is not a surprising result, as many authors have concluded that
the Abell catalog becomes incomplete at $z\gtapr0.1$ (e.g., Southerland 1988;
Struble \& Rood 1991b).

\subsection {A \bgc~Calibrated Richness Classification}

Using Figures 4 or 5, we can relate the more robust \bgc~measurements
directly to the Abell Richness class, in essence, calibrating the
classical richness class designations.
The mediam \bgc~value of ARC 1 clusters in our sample is very close
to 1000 \mpc17.
Hence, we propose to use 1000 \mpc17~as the fiducial \bgc~value for
the average ARC 1 counts.

Abell (1958) and ACO used a non-linear relationship between the Abell
counts and the richness class with the higher ARC having disproportionately
higher counts, presumably to account for the few apparently very high
$N_A$ clusters . 
However, \S 5.2 demonstrates that  the very high $N_A$ clusters
 are probably much less rich.
If we scale the \bgc~values corresponding to each ARC counts using
the fiducial of 1000 \mpc17~as ARC 1, we see that there is a significant
mismatch at ARC$\ge$3, as illustrated by the dotted line in Figure 4,
with ARC 5 being equivalent to \bgc$\sim7000$.
Yee et al. (1999) also found that none of the 16 clusters from the
CNOC1 survey, which contains the most X-ray luminous clusters
from the EMSS survey (see Gioia \& Luppino 1994), have \bgc~larger than
$\sim 2500$ Mpc $^{-1.8}$.
Hence, there appears to be no known clusters that approach the original
definition of ARC 5.
In view of this, we propose a linear richness scale where adjacent richness
classes change by the same increment of 400 \mpc17~in \bgc.
This scale would cover most of the range of known cluster richness
(including very rich, X-ray luminous,
 higher redshift clusters, such as those
in the CNOC1 sample),
with the first 3 classes (0, 1, and 2) matching the original ARC
definition reasonably well.
This revised classification definition is shown graphically in Figure 4.
For $H_0=50$ km s$^{-1}$ Mpc $^{-1}$, the calibration for ARC 0 to 5 
is then: 600$\pm$200, 1000$\pm$200, 1400$\pm$200, 1800$\pm$200, 
2200$\pm$200, and $>2400$ \mpc17, respectively.

We note that this calibration is substantially different from those
provided by Prestage \& Peacock (1988) based on a very small number
of 4C radio galaxies situated in Abell clusters.
Their calibration of ARC 0 and 1 having \bgc~of $\sim$380 and 850 \mpc17~is
about 1/3 to 1/2 of an Abell class too high.
Hence, previous descriptions of the Abell class equivalence for the
environments of radio galaxies and quasars (e.g., YG87) need to be
revised correspondingly downward.

\subsection {The Correlation Between Velocity Dispersion and Richness}

So far we have shown that a richness parameter such as \bgc, which
uses photometric data and normalized by the LF and spatial
distribution of galaxies, produces a robust measurement of the
richness of a cluster.
One would then expect that such a robust parameter will provide a
better description of the clusters, and hence produce more
meaningful correlations with other cluster properties.

We will show, as a demonstration of the improvement expected, an
example of such a correlation.
One fundamental cluster property is the mass of a cluster, which
can be estimated by the velocity dispersion ($\sigma$).
In Figure 6 we plot the correlation of $\sigma$ versus \bgc.
The velocity dispersion measurements are taken from the compilation
of Struble \& Rood (1991a), and Fadda et al. (1996).
We include only those clusters with 10 or more velocity measurements.
There is a clear correlation between $\sigma$ and \bgc, and
the correlation is much tighter than using $N_A$ as the variable
for richness which is shown in Figure 7.
Girardi et al. (1993) showed a similar improvement in the
correlation between velocity dispersion and richness when
Bahcall's (1981) $N_{0.5}$ parameter is used instead of the Abell counts.
This improved correlation lends support that \bgc~is a more
robust richness measurement.

The $\sigma$-\bgc~correlation derived from the CNOC1 cluster
redshift survey (Yee et al. 1999) provides an even tighter
correlation between the two quantities. 
This is almost certainly due to the fact that the velocity
dispersion measurements from CNOC1 are derived from a 
homogeneous set of velocity data with a large number of
velocities for each cluster.
We note that the $\sigma$-\bgc~relationships derived from the two
different samples (this work and CNOC1) are identical within the 
uncertainty.
The $\sigma$--\bgc~relationship from CNOC1 is plotted as a dotted
line in Figure 6.
Given the different photometric systems used (Kron-Cousin $R$
vs Gunn $r$), the different redshift range of the sample, and
the different instrumentations used for these two samples,
this again points to the robustness of the \bgc~measurements
as a richness estimator.

\section{Summary}

We have used
a sample of 47 Abell clusters with photometric catalogs from 
CCD images to
test the robustness of \bgc~as a parameter for measuring the richness
of galaxy clusters.
We first tested the effect of the choice of galaxy LF, which is used
to normalize the excess galaxy counts.
Using three LFs derived from different sources, including both cluster
and field galaxies, we found that as long as we adjust the density
 normalization
constant of the LF so that it reproduces the background counts used
to derive the excess counts, different choices of the LF produce 
variations in \bgc~of approximately 10\%.
Furthermore, the parameters for the LF can be substantially incorrect
(e.g., with $M^*$ being off by 0.5 mag), errors in \bgc~can also
be minimized by renormalizing the LF to fit the background counts.

We also found that if a correct LF is used, counting to different absolute
magnitude limits produces systematic variations of only a few percent.
However, to minimize uncertainties introduced by LF discrepancy and
photometry errors, the results are most reliable when the counting is
done to about 1 to 2 mag past $M^*$.

Similarly, the canonical $\gamma=1.8$ for the spatial correlation function of
field galaxies provides  reasonably stable \bgc~values as one varies
the radius of the counting aperture.
Altering the counting radius from 0.25 to  1.5 Mpc produces systematic
changes in \bgc~of less than 10\%.

We have compared our \bgc~richness parameters with the Abell Richness Class
and the Abell Count Number ($N_A$).
Limiting the sample to either ARC$\le$2 or $z\ltapr0.1$, we found 
a good correlation between $N_A$ and \bgc.
However, clusters designated as ARC$\ge$3 appear often to have 
their richness over estimated, sometimes by as much as a factor of 3.
Since most of the ARC$\ge$3 clusters are at $z>0.1$,
this suggests that ARC classifications for
clusters with $z\gtapr0.1$ is not a good indicator of their true richness.
We suggest a revised richness classification which is
 based on a linear scale in \bgc.
  Finally, we demonstrate the improvement in the correlation between
richness and velocity dispersion when a robust richness parameter,
like \bgc, is used.

\acknowledgements

We wish to thank KPNO for the generous allotment of 
telescope time and assistance in carrying out the
Low-redshift Cluster Optical Survey (LOCOS).  HY acknowledges NSERC
for financial support in the form of an operating grant.  OLC was
supported in part by an overseas scholarship by CONACyT-M\'exico and
INAOE.  OLC also acknowledges financial support from the Department of
Astronomy and the School of Graudate Studies of the University of
Toronto, NSERC through HY's operating grant, and CONACyT-M\'exico through a
``Catedra de Repatriaci\'on'' and  a ``Proyecto de Investigaci\'on Inicial.''

\bigskip
\bigskip

\newpage

\figcaption{
Background galaxy counts in Cousin $R$ from 9 KPNO 0.9m
fields.  The dashed, dotted, and solid lines represent 
the count models created using the LCY LF, KE LF, and CNOC2 LF,
respectively.
}   
\figcaption{
Comparison of \bgc~values computed using counting 
magnitude limits of --20.0 and --21.0.
The former cut-off is equivalent to $M^*-1.7$.
The dotted line is not a fit, but a line of slope unity.
}   
\figcaption{
Comparison of \bgc~values computed using counting 
radii of 0.5 Mpc and 1.0 Mpc.
The dotted line is not a fit, but a line of slope unity.
}
\figcaption{
The correlation between Abell Richness Class and the \bgc~parameter.
The  median \bgc~value for each ARC is shown as an asterix.
The dot-dashed line indicates the expected \bgc~values for each
ARC which describes 
the relative median richness of ARC 1 and 2 well, but fails significantly
for richer classes.
The six bars connected by the dashed line illustrates a linear
richness classification which matches the real richness of the
various ARC reasonably well.
}
\figcaption{
The correlation between Abell count number and the \bgc~parameter.
Solid symbols represent cluster with $z<0.09$; while open symbols
mark clusters with $z>0.09$.
The solid star symbols are Abell clusters at $\sim 0.2$ from the CNOC1 survey.
The dashed line shows the best fitting line for the $z<0.09$ subsample
force-fitted through the origin, while the  dotted line is the best
linear fit for the $z>0.09$ subsample.
}
\figcaption{
The correlation between velocity dispersion and cluster
richness as parametrized by \bgc.
The dashed line is the best linear fit; while
the dotted line shows the relationship derived from 15 CNOC1 clusters
(Yee et al. 1999).
}
\figcaption{
The much poorer correlation between velocity dispersion and cluster
richness as parametrized by the Abell count number ($N_A$).
}

\newpage


\begin{deluxetable}{lccrc} 
\tablenum{1} 
\tablewidth{3.5 true in} 
\tablecaption{Abell Clusters} 
\tablehead{ 
\colhead{Cluster} & 
\colhead{$z$} & 
\colhead{ARC} & 
\colhead{$B_{gc}$} & 
\colhead{$\pm$}  
} 

\startdata 

  A21   &    0.0946 & 1 &  1509 &  241 \nl
  A84   &    0.1030 & 1 &  1000 &  204 \nl
  A85   &    0.0518 & 1 &   757 &  174 \nl
  A98   &    0.1043 & 3 &  1921 &  268 \nl
  A154  &    0.0638 & 1 &  1393 &  230 \nl
  A168  &    0.0452 & 2 &   973 &  192 \nl
  A399  &    0.0715 & 1 &  1449 &  235 \nl
  A401  &    0.0748 & 2 &  2242 &  286 \nl
  A407  &    0.0472 & 0 &  1368 &  225 \nl
  A415  &    0.0788 & 1 &   534 &  159 \nl
  A514  &    0.0731 & 1 &   946 &  196 \nl
  A629  &    0.1380 & 1 &  1205 &  224 \nl
  A646  &    0.1303 & 0 &   858 &  196 \nl
  A665  &    0.1816 & 5 &  2186 &  290 \nl
  A671  &    0.0491 & 0 &  1266 &  217 \nl
  A690  &    0.0788 & 1 &   701 &  175 \nl
  A957  &    0.0437 & 1 &  1075 &  201 \nl
  A1213 &    0.0469 & 1 &   931 &  189 \nl
  A1291 &    0.0530 & 1 &  1122 &  207 \nl
  A1413 &    0.1427 & 3 &   1686 &  257 \nl
  A1569 &    0.0784 & 0 &    768 &  181 \nl
  A1650 &    0.0845 & 2 &   1861 &  263 \nl
  A1656 &    0.0232 & 2 &   1242 &  282 \nl
  A1775 &    0.0700 & 2 &   1018 &  202 \nl
  A1795 &    0.0621 & 2 &   1430 &  233 \nl
  A1913 &    0.0530 & 1 &    954 &  192 \nl
  A1983 &    0.0430 & 1 &   974 &  192 \nl
  A2029 &    0.0768 & 2 &  1736 &  255 \nl
  A2244 &    0.0997 & 2 &  1698 &  254 \nl
  A2255 &    0.0800 & 2 &  2296 &  289 \nl
  A2256 &    0.0601 & 2 &  2174 &  281 \nl
  A2271 &    0.0568 & 0 &   645 &  164 \nl
  A2328 &    0.1470 & 2 &  1941 &  273 \nl
  A2356 &    0.1161 & 2 &   944 &  201 \nl
  A2384 &    0.0943 & 1 &  1575 &  245 \nl
  A2399 &    0.0587 & 1 &   906 &  190 \nl
  A2410 &    0.0806 & 1 &   699 &  175 \nl
  A2415 &    0.0597 & 0 &   902 &  190 \nl
  A2420 &    0.0838 & 2 &  1230 &  220 \nl
  A2440 &    0.0904 & 0 &  1053 &  207 \nl
  A2554 &    0.1108 & 3 &  1218 &  222 \nl
  A2556 &    0.0865 & 0 &   828 &  187 \nl
  A2593 &    0.0421 & 0 &  1211 &  212 \nl
  A2597 &    0.0825 & 0 &   665 &  172 \nl
  A2626 &    0.0573 & 0 &   945 &  193 \nl
  A2657 &    0.0414 & 1 &   740 &  170 \nl
  A2670 &    0.0761 & 3 &  1771 &  257 \nl
\enddata
\end{deluxetable} 



\begin{deluxetable}{llllrrrr} 
\tablenum{2} 
\tablewidth{6 true in} 
\tablecaption{Luminosity Functions} 
\tablehead{ 
\colhead{LF} & 
\colhead{galaxy type} & 
\colhead{$\phi'$} & 
\colhead{$\phi^*$}  &
\colhead{$\alpha$}  &
\colhead{$M^*$}  &
\colhead{$Q$}  &
\colhead{$P$}  
} 

\startdata 

 LCY   & cluster & 0.00226  &  1.0 & --1.00 & --22.20 &  1.4  & 0.0 \nl
       &    &          &      &      &       &       &     \nl
 KE    & E/S0 &  0.000346  &  1.79 & --1.00 & --22.45 &  1.4  & 0.0 \nl
       & Sab  &  0.000346  &  1.87 & --1.00 & --22.14 &  1.4  & 0.0 \nl
       & Sbc  &  0.000346  &  1.87 & --1.00 & --21.81 &  1.4  & 0.0 \nl
       & Scd/Sdm   &  0.000346  &  3.24 & --1.00 & --21.29 &  0.7  & 0.0 \nl
       &    &          &      &      &       &       &     \nl
 CNOC2 & early        &  0.92     &  0.00212  &  0.09 & --21.52 &  1.94 & --2.02 \nl
       & intermediate &  0.92     &  0.00111  & --0.69 & --21.84 &  0.62 & 0.59 \nl
       & late         &  0.92     &  0.00053  & --1.45 & --21.44 &  0.88 & 2.68 \nl

\enddata
\end{deluxetable} 



\begin{deluxetable}{cclrrrrr} 
\tablenum{3} 
\tablewidth{6 true in} 
\tablecaption{Fractional Change to mean $B_{gc}$ Adjusting $M^*$ } 
\tablehead{ 
\colhead{Sampling Limit} &
\colhead{LF Renormalization} & 
\colhead{$\Delta M^*$:} & 
\colhead{--0.50} & 
\colhead{--0.25} & 
\colhead{0.00}  &
\colhead{+0.25}  &
\colhead{+0.50} 
} 

\startdata 

--20 & yes & &  1.24 & 1.11 & 1.00 & 0.91 & 0.84 \nl
--20 & no  & & 0.80 & 0.89 & 1.00 & 1.14 & 1.32 \nl
--22 & yes & & 0.90 & 0.95 & 1.03 & 1.18 & 1.42 \nl
--22 & no & & 0.58 & 0.76 & 1.03 & 1.47 & 2.24 \nl

\enddata
\end{deluxetable}



\begin{deluxetable}{cclrrrrr} 
\tablenum{4} 
\tablewidth{6 true in} 
\tablecaption{Fractional Change to mean $B_{gc}$ Adjusting $\alpha$ } 
\tablehead{ 
\colhead{Sampling Limit} &
\colhead{LF Renormalization} & 
\colhead{$\Delta \alpha$:} & 
\colhead{--0.30} & 
\colhead{--0.15} & 
\colhead{0.00}  &
\colhead{+0.15}  &
\colhead{+0.30} 
} 

\startdata 

--20 & yes & &  0.92 & 0.96 & 1.00 & 1.10 & 1.19 \nl
--20 & no  & & 0.73 & 0.86 & 1.00 & 1.14 & 1.29 \nl
--22 & yes & & 1.39 & 1.19 & 1.03 & 0.95 & 0.87 \nl
--22 & no & & 1.11 & 1.07 & 1.03 & 0.99 & 0.94 \nl

\enddata
\end{deluxetable} 

\newpage



\begin{figure}[h] \figurenum{1}\plotone{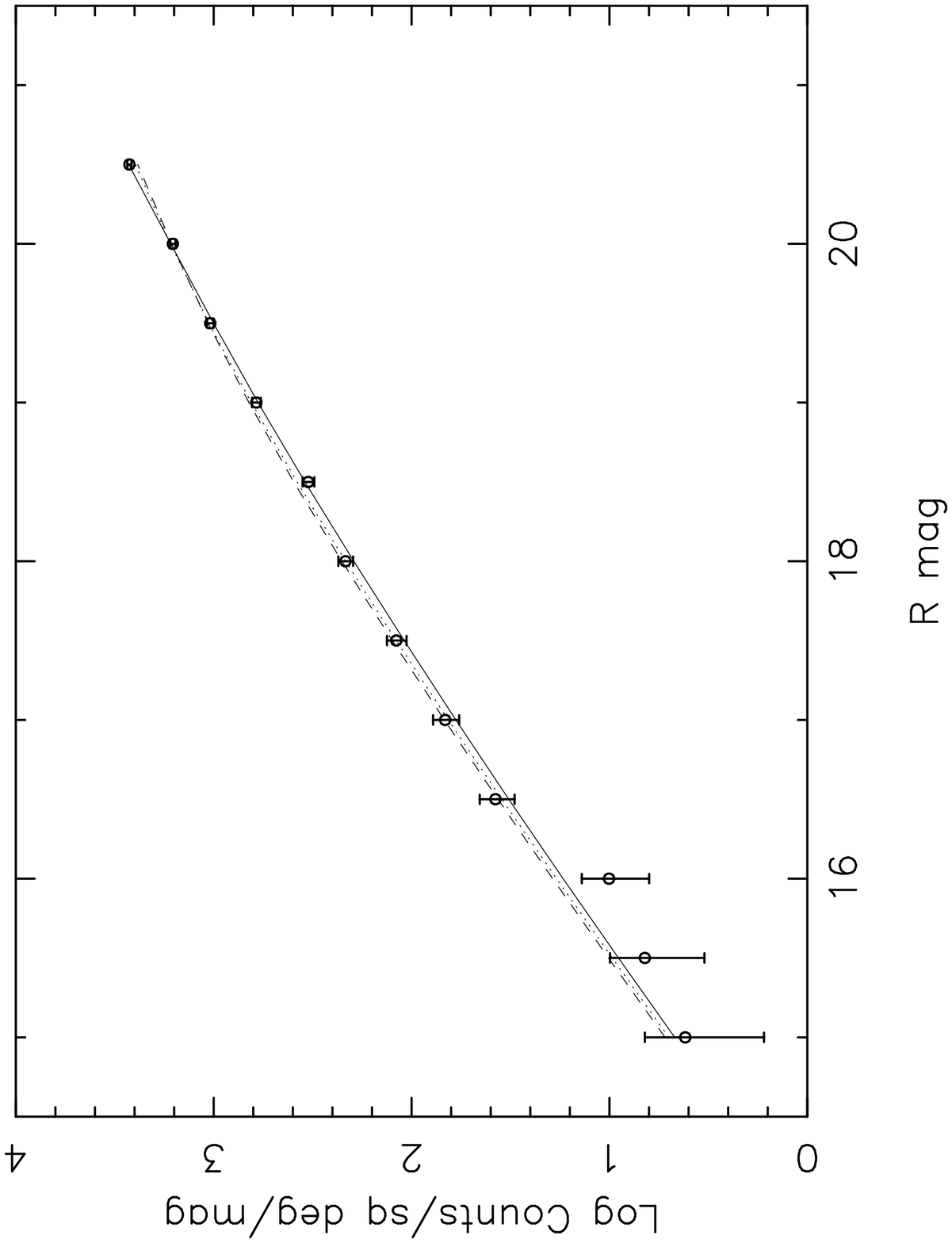} \caption{}\end{figure}
\begin{figure}[h] \figurenum{2}\plotone{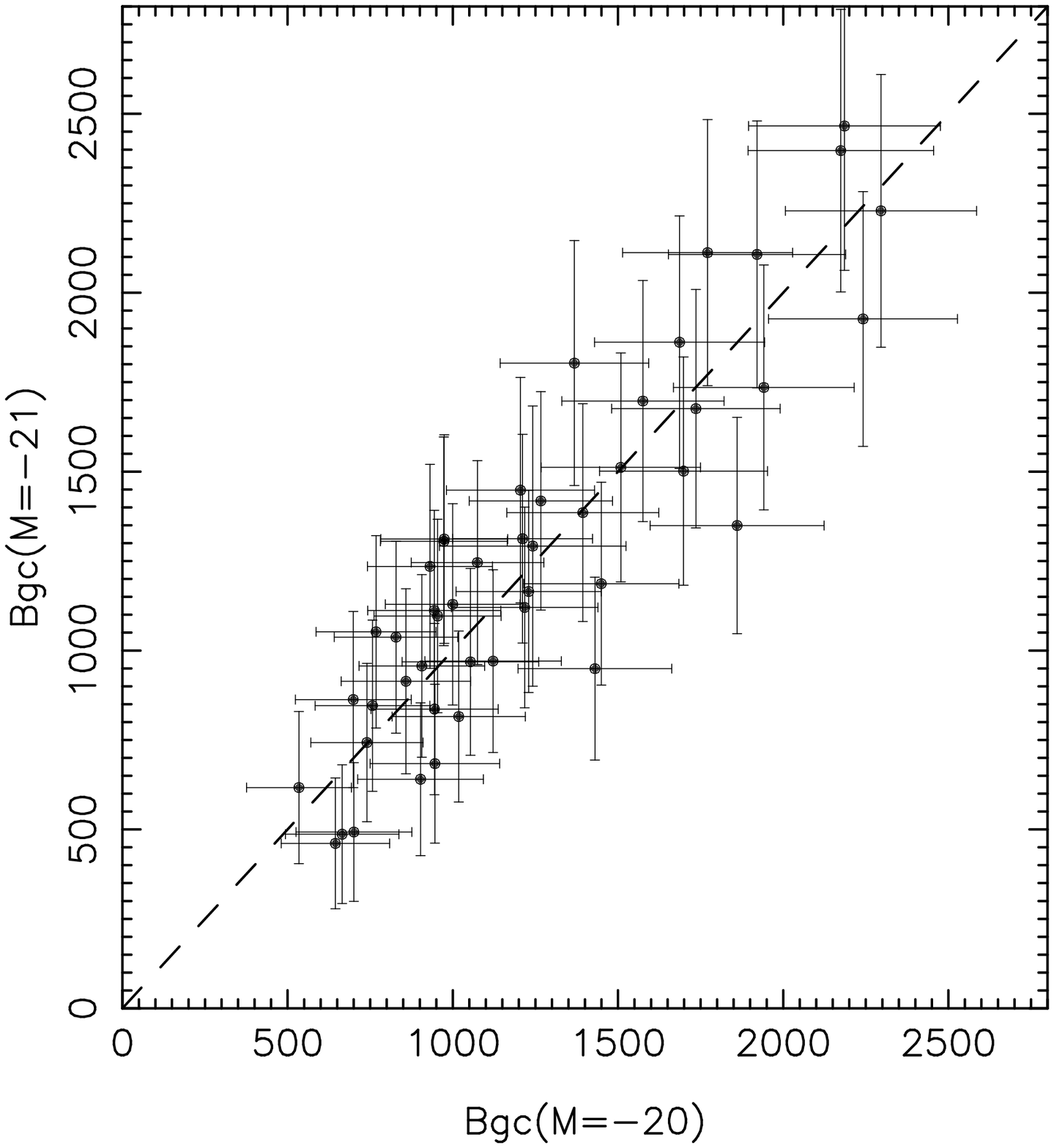} \caption{}\end{figure}
\begin{figure}[h] \figurenum{3}\plotone{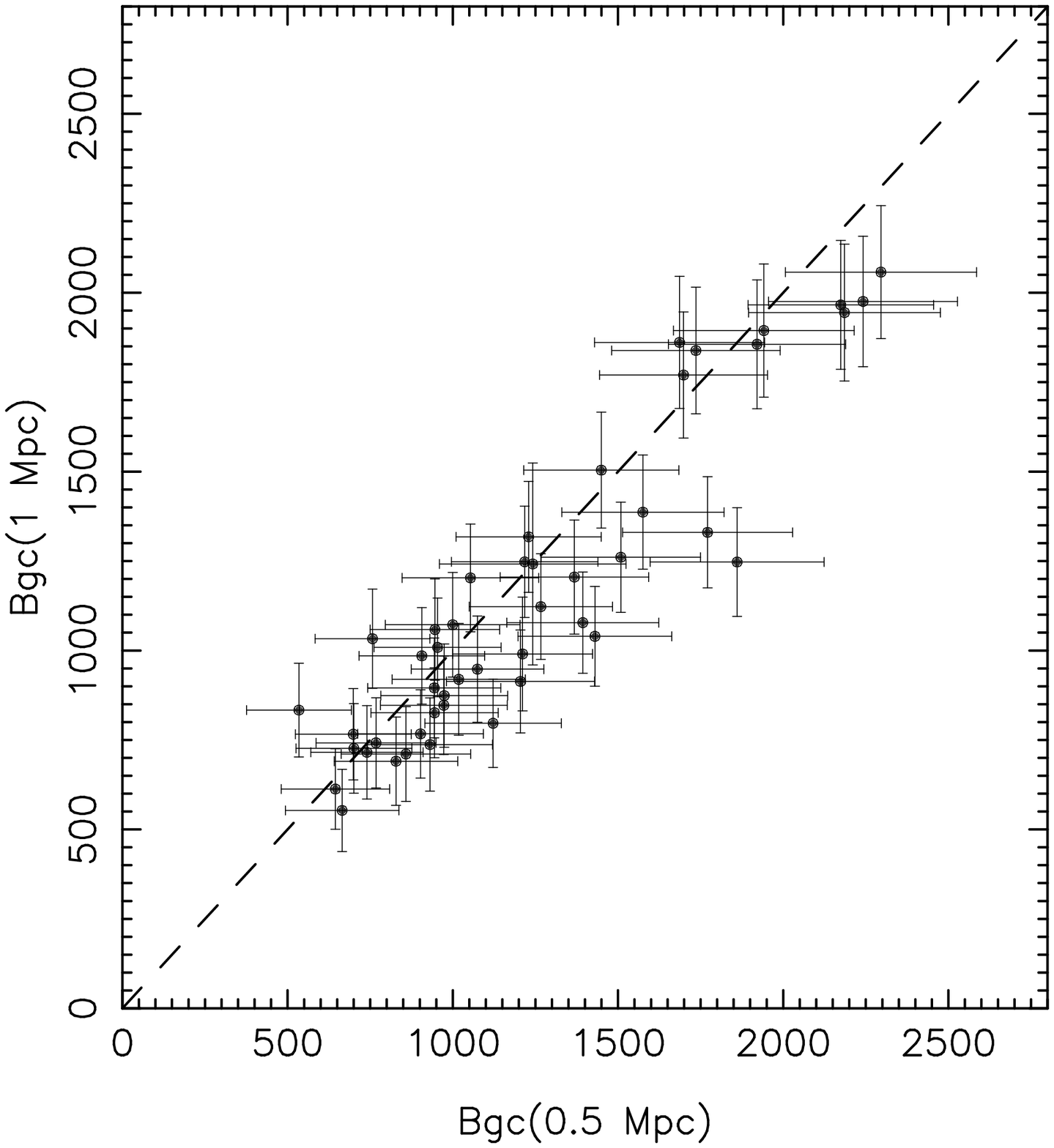} \caption{}\end{figure}
\begin{figure}[h] \figurenum{4}\plotone{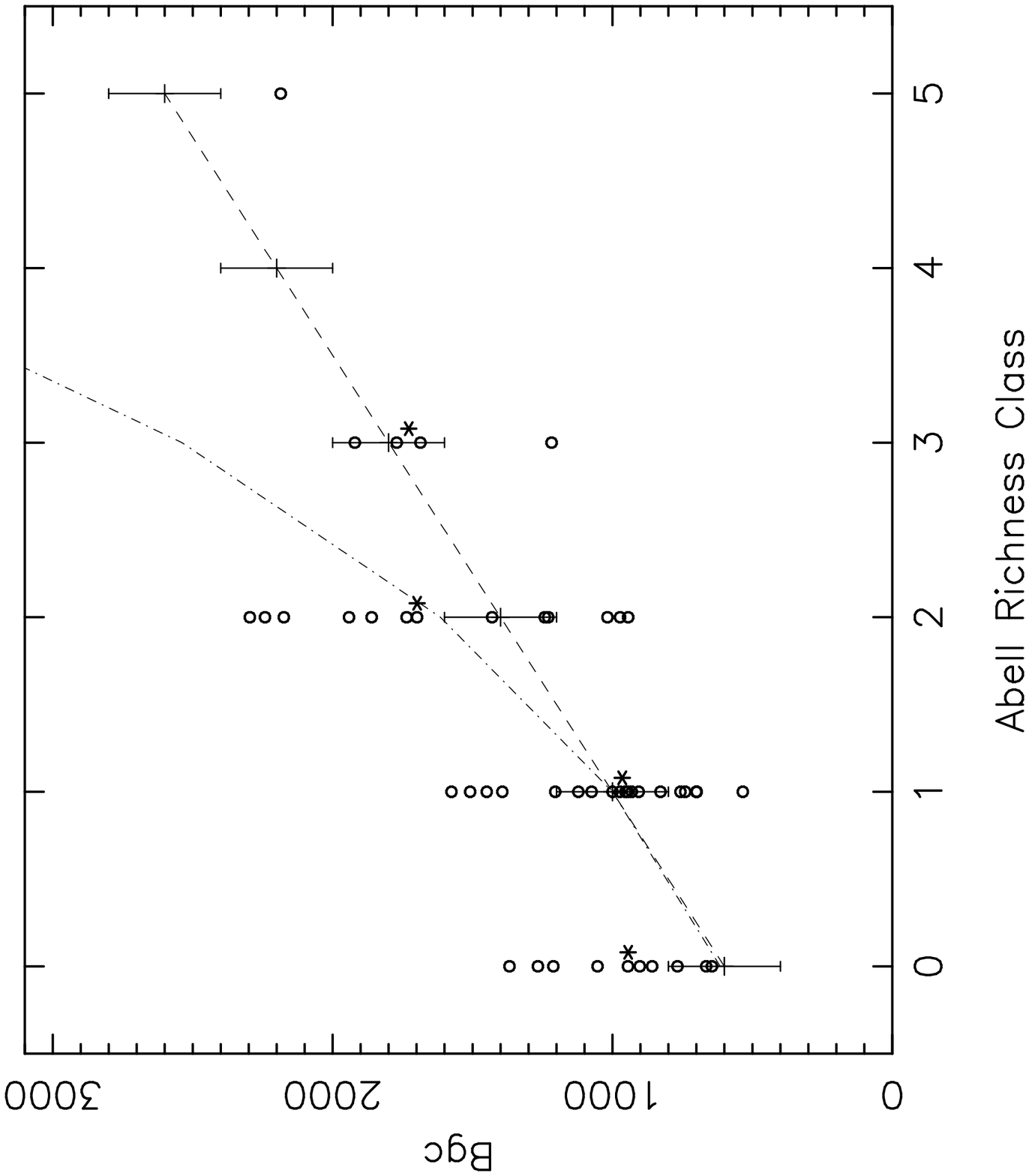} \caption{}\end{figure}
\begin{figure}[h] \figurenum{5}\plotone{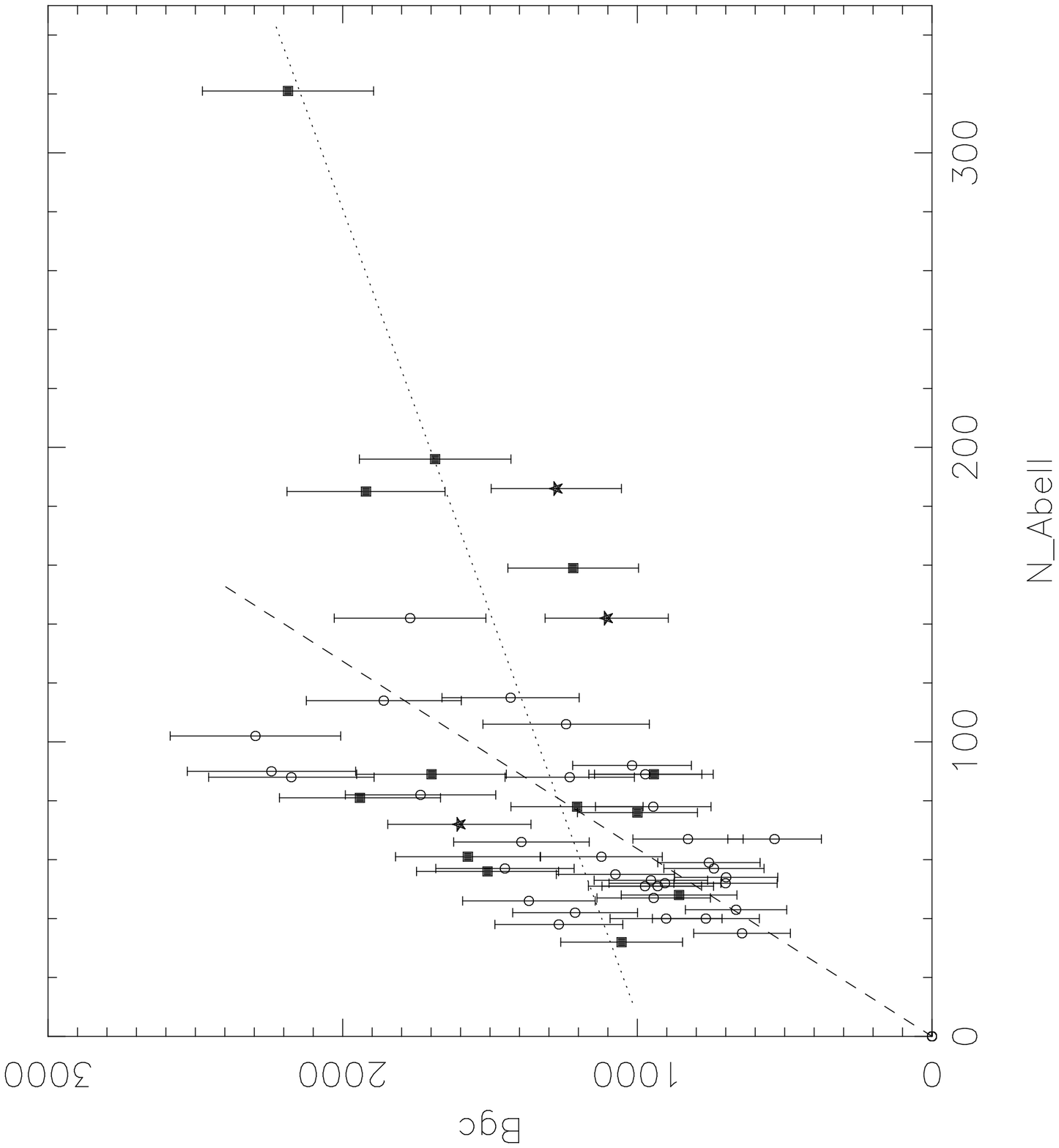} \caption{}\end{figure}
\begin{figure}[h] \figurenum{6}\plotone{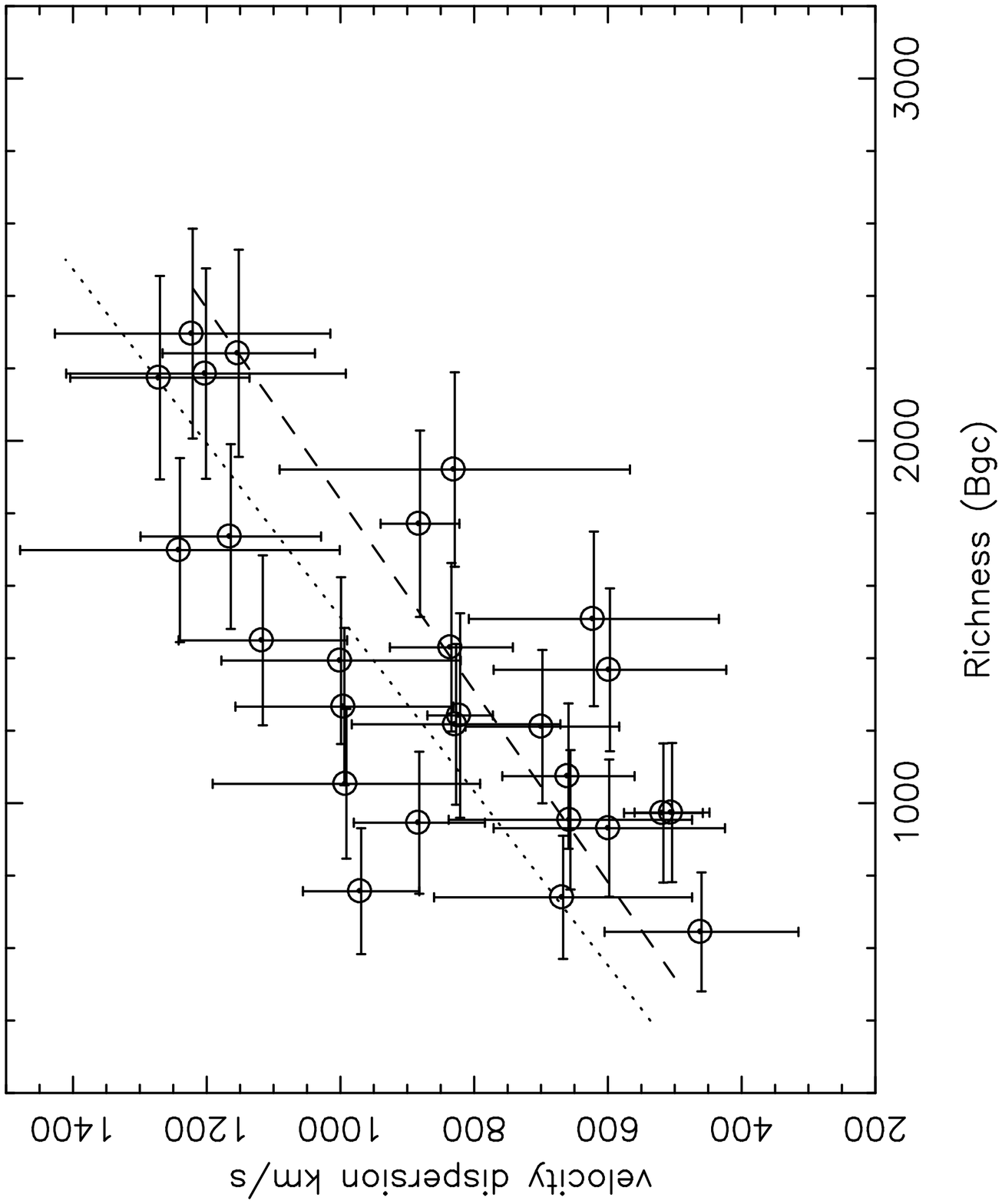} \caption{}\end{figure}
\begin{figure}[h] \figurenum{7}\plotone{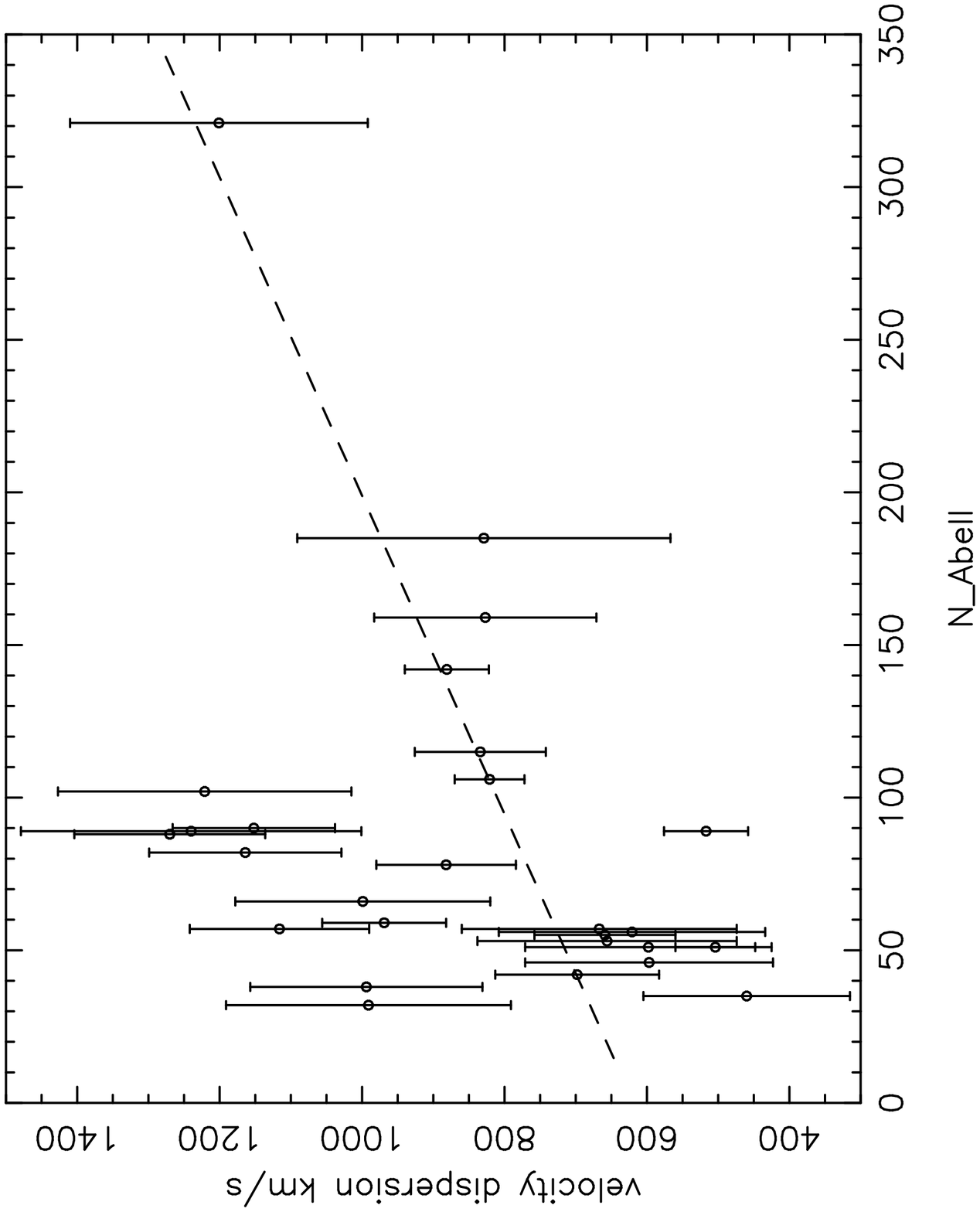} \caption{}\end{figure}

\end{document}